\documentclass[fleqn,twoside]{article}

\usepackage{espcrc2}
\usepackage{graphicx}
\usepackage{amssymb}

\newcommand{\bc}{\begin{center}}
\newcommand{\ec}{\end{center}}
\newcommand{\be}{\begin{equation}}
\newcommand{\ee}{\end{equation}}
\newcommand{\bea}{\begin{eqnarray}}
\newcommand{\eea}{\end{eqnarray}}
\newcommand{\bi}{\begin{itemize}}
\newcommand{\ei}{\end{itemize}}
\newcommand{\bd}[1]{\begin{dinglist}{#1}}
\newcommand{\ed}{\end{dinglist}}
\newcommand{\bt}{\begin{tabular}}
\newcommand{\et}{\end{tabular}}

\def\csw{c_{\rm SW}}

\def\ksea{\kappa^{\rm sea}}
\def\mq{m_{\rm q}}

\def\qt{Q_{\rm t}}
\def\chit{\chi_{\rm t}}

\title{%
      \vspace{-1.00cm}					
      {\normalsize DESY 02--173} \\			
      {\normalsize Edinburgh 2002/07} \\		
      \vspace{+1.00cm}					
      Low-Lying Fermion Modes: Dynamical versus Quenched%
      \thanks{Poster presented by D.~Pleiter at the Lattice Conference 2002,
              Cambridge (MA), USA.}
}

\author{%
  {\it QCDSF} and {\it UKQCD} Collaboration:
  R.~Horsley\address[Ed]{%
	School of Physics,
	University of Edinburgh,
	Edinburgh EH9 3JZ, UK},
  T.G.~Kov\'{a}cs\address[Ztn]{%
	John von Neumann Institute NIC / DESY Zeuthen,
	D-15738 Zeuthen, Germany},
  V.~Linke\address[FU]{%
	Institut f\"{u}r Theoretische Physik,
	Freie Universit\"{a}t Berlin,
	D-14195 Berlin, Germany},
  D.~Pleiter\addressmark[Ztn],
  G.~Schierholz\addressmark[Ztn]$^{,}$\address[HH]{%
	Deutsches Elektronen-Synchrotron
	DESY, D-22603 Hamburg, Germany},
  T.~Streuer\addressmark[Ztn]$^{,}$\addressmark[FU]
}

\begin{document}

\begin{abstract}
We compare the low-lying eigenmodes of the $O(a)$ improved Wilson-Dirac
operator on quenched and dynamical configurations
and investigate methods of probing the topological properties of
gauge configurations.
\vspace*{-0.2cm}
\end{abstract}

\maketitle

\section{INTRODUCTION}

The eigenmodes of the Dirac operator carry information about the
topological content of the background gauge field. The Atiyah-Singer index
theorem tells us that in a sector with topological charge $\qt$
the Dirac operator has at least
$|\qt|$ zero modes. In the presence
of light dynamical fermions, the light quark determinant should
suppress topological sectors with large $|\qt|$.
As a consequence we expect a reduced value for the topological
susceptibility
\be
\chit = \frac{\langle\qt^2\rangle}{V}.
\ee

The topological charge is still difficult to determine on the lattice.
Various methods have been explored, usually gluonic methods.
The latter, however, typically require smoothing procedures
to reduce the ultraviolet fluctuations which do not take the
fermionic part of the action, i.e. the presence of dynamical
fermions, into account.

Fermionic methods, on the other hand, used to probe topological
properties are expected to be sensitive to the effects caused by
explicitly breaking chiral symmetry when using Wilson fermions.
These effects are potentially large on coarse lattices, even when
one reduces these effects using $O(a)$ Symanzik improved fermions.
\vspace*{-0.6cm}
\begin{figure}[h]
\includegraphics[scale=0.5]{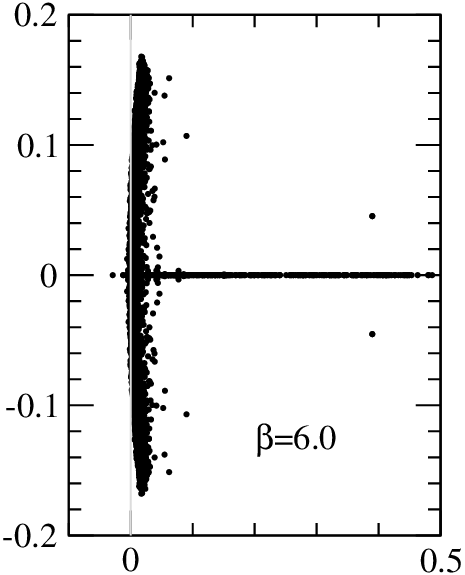}\hfill
\includegraphics[scale=0.5]{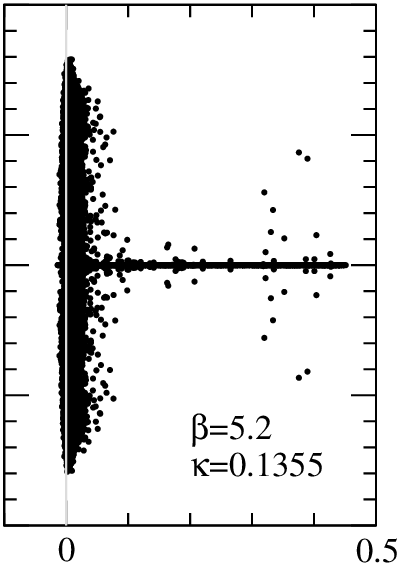}
\vspace*{-1cm}
\caption{Eigenvalues of the massless improved Wilson-Dirac operator calculated
on $O(150)$ quenched configurations (left) and dynamical (right)
configurations.}
\vspace*{-1.0cm}
\end{figure}

\section{THE NON-HERMITIAN CASE}

The $O(a)$ improved Wilson-Dirac operator
\be
M(\kappa) =
1 - \kappa\,\left[H + \frac{i}{2}\,\csw F_{\mu\nu}\sigma_{\mu\nu}\right],
\ee
is a non-hermitian operator. To calculate its eigenvalues $\lambda_i$
we used the Arnoldi algorithm \cite{arpack}.  To speed up convergence
and to maximize the number of real modes found by the algorithm, we
transformed $M$ using a ``least-squares'' polynomial of degree $40$
\cite{saad}. This polynomial was tuned to make the eigenvalues with small
real part lying in a band along the real axis fastest to converge.
For our purpose it is crucial to increase the number of calculated small
real modes.
We calculated $O(100)$ eigenvalues per configuration.

\begin{figure}[t]
\vspace*{0.1cm}
\hspace*{1.0mm}\includegraphics[scale=0.5]{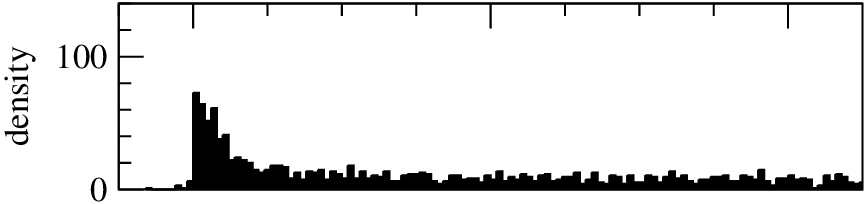} \\
\hspace*{0.0mm}\includegraphics[scale=0.5]{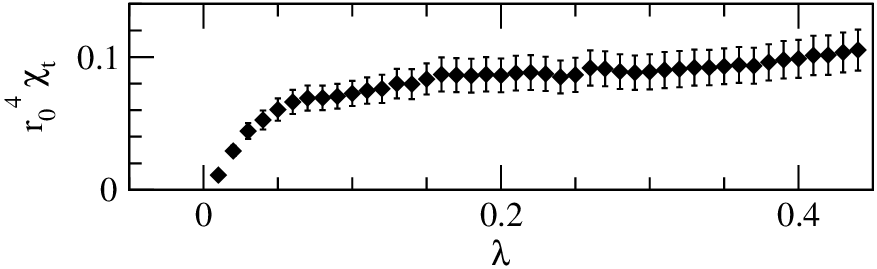}
\vspace*{-1.5cm}
\caption{The density of real modes (upper plot) and the topological
susceptibility as a function of the largest real mode taken into
account (lower plot) calculated on quenched configurations
at $\beta=5.9$.\label{fig:dens_susc_b5p90}}
\vspace*{-0.5cm}
\end{figure}

\begin{figure}[t]
\vspace*{0.05cm}
\hspace*{1.0mm}\includegraphics[scale=0.5]{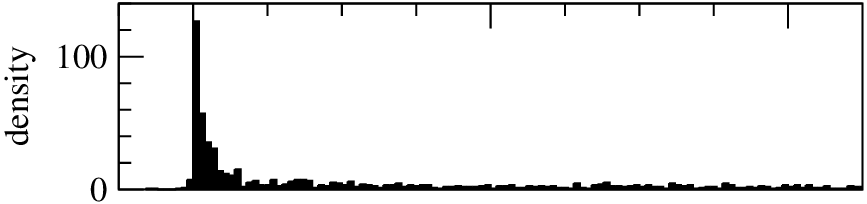} \\
\hspace*{0.0mm}\includegraphics[scale=0.5]{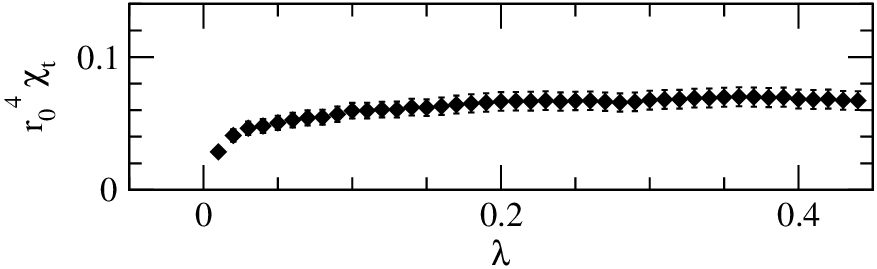}
\vspace*{-1.5cm}
\caption{Same as in Fig.~\ref{fig:dens_susc_b5p90} for quenched configurations
at $\beta=6.0$.\label{fig:dens_susc_b6p00}}
\vspace*{-0.5cm}
\end{figure}

\begin{figure}[ht]
\hspace*{+0.4mm}\includegraphics[scale=0.5]{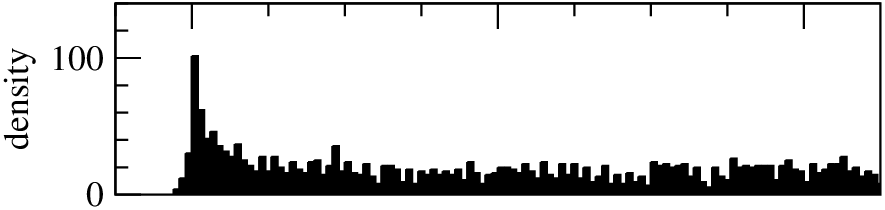} \\
\hspace*{-0.5mm}\includegraphics[scale=0.5]{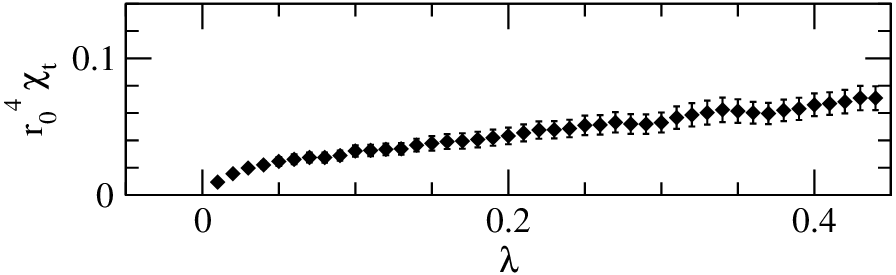}
\vspace*{-1.5cm}
\caption{Same as in Fig.~\ref{fig:dens_susc_b5p90} for dynamical configurations
at $(\beta,\ksea) = (5.2, 0.1355)$.\label{fig:dens_susc_b5p20kp13550}}
\vspace*{-0.8cm}
\end{figure}

The index theorem establishes a relation between the difference in the
number of real modes with negative ($n^+$) and positive ($n^-$) chirality
$\omega_i = (v_i, \gamma_5 v_i)$ and the topological charge:
\be
\qt = n^+ - n^-.
\label{eq:qt_index}
\ee
However, on the lattice this relation can only hold approximately
(if at all) in case of Wilson fermions. We can assume
this approximation only to be reasonable if the spectrum of the
Wilson-Dirac operator allows us to distinguish between physical modes and
unphysical modes due to the doublers.
To check for this precondition we considered the density of real modes.
If our massless Dirac operator would be con\-ti\-nuum-like we would
expect the density of real modes being sharply peaked around zero and
becoming small for larger modes. Comparing the densities at two different
values of the gauge field coupling $\beta$ in the quenched case we indeed
see a more continuum-like behaviour for larger $\beta$. In case of the
dynamical configurations we expect small eigenvalues to be suppressed
and therefore the density of real modes around zero to be smaller. This
is indeed what we find, see
Figs.~\ref{fig:dens_susc_b5p90}, \ref{fig:dens_susc_b6p00}
and \ref{fig:dens_susc_b5p20kp13550}.
We now consider $\chit$ using $\qt$ from Eq.~(\ref{eq:qt_index}) taking
all real modes $\lambda_i < \lambda$ into account.
For larger values of the cut-off $\lambda$ one would expect $\chit$
to become constant. In the quenched case $\chit$ shows a reasonable
plateau, which seems to improve for larger $\beta$.
The same analysis done on an ensemble of dynamical configurations
does not show any plateau.

\section{THE HERMITIAN CASE}

The Wilson-Dirac operator becomes Hermitian when multiplied by $\gamma_5$.
To calculate $O(100)$ smallest eigenvalues of $\gamma_5 M(\kappa)$
we used again the Arnoldi algorithm with a Chebychev polynomial to
speed up convergence of the eigenvalues with smallest modulus. Note
that there is no trivial relation between the eigenvalues of $\gamma_5
M(\kappa)$ for different values of $\kappa$, while
the eigenvalues of $M(\kappa)$ for different $\kappa$ are related by a
trivial scale and shift operation. We used $\kappa = 0.1338$ and $0.1353$
for $\beta = 6.0$ and $(\beta,\ksea) = (5.2, 0.1355)$,
respectively, which corresponds to $m_{\rm PS}/m_{\rm V} \approx 0.6-0.7$.

\begin{table}[t]
\caption{Simulation parameters. We used lattices of size $V=16^3\times 32$.
The scale was set using $r_0 = 0.5$fm.}
\bt{l|l|l|l}
$\beta$ & $\ksea$, $m_{\rm PS,sea}$ [MeV] & $a$ [fm] & \#conf  \\
\hline
5.9     & quenched                        & 0.112    & $O(100)$ \\
6.0     & quenched                        & 0.093    & $O(150)$ \\
5.2     & 0.13550, 578(6)                 & 0.099    & $O(150)$ \\
\et
\vspace*{-0.5cm}
\end{table}

The eigenvalues of the Hermitian operator can be used to calculate the
topological charge $\qt$. Based on a chiral Ward identity one can define
\be
\qt = -\lim_{\mq \rightarrow 0} \mq {\rm Tr}\left[\gamma_5 M(\mq)^{-1}\right].
\label{eq:qt_cwi}
\ee

If the largest of the $N$ calculated smallest eigenvalues is large enough,
we expect
\be
{\rm Tr}\left[\gamma_5 M(\mq)^{-1}\right] \approx
\sum_{i=1}^N \frac{1}{\lambda_i},
\label{eq:trq}
\ee
to be a good approximation. For an extrapolation
to $N=\dim(\gamma_5 M)$ we performed a fit using the ansatz
$c_0 + c_{-1}\,|\lambda|^{-1}$ \cite{neff}.

\begin{figure}
\bc
\includegraphics[scale=0.5]{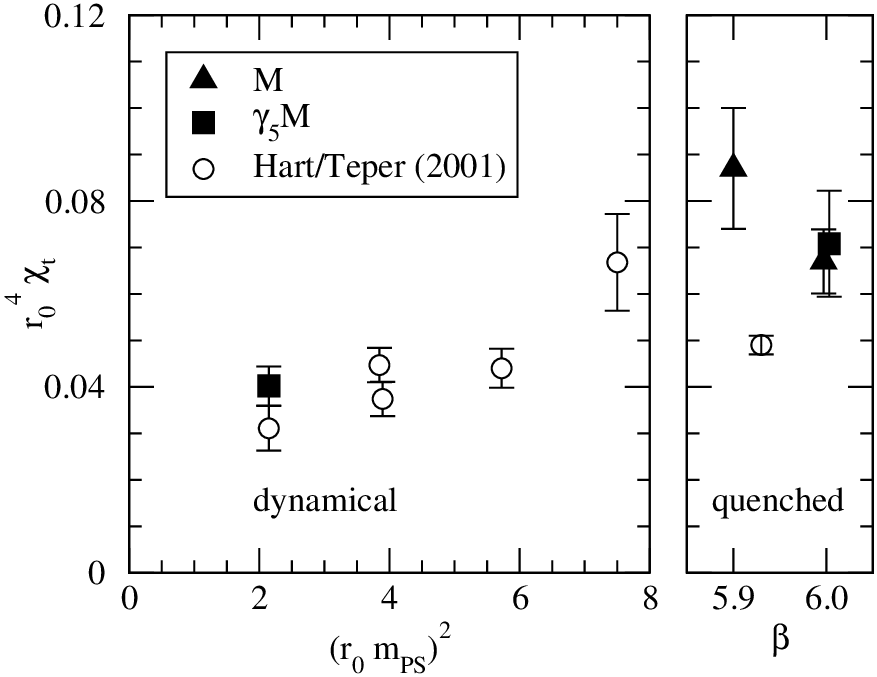}
\ec
\vspace*{-1.4cm}
\caption{Results for the topological susceptibility using different
fermionic methods (solid symbols, this work) and a gluonic method
(open symbols, \cite{Hart:2001fp}).\label{fig:suscept}}
\vspace*{-0.8cm}
\end{figure}

While the theoretical basis for the definition of the topological
charge using Eq.~(\ref{eq:qt_cwi}) is much better, there are a
number of disadvantages and open problems. For instance,
calculations have to be performed at quark masses which are heavy enough
to avoid effects from ``exceptional configurations''. Therefore,
a chiral extrapolation is in principle required, but has not been done
here. Initial results using smaller (and coarser) lattices show
a very mild quark mass
dependence \cite{tamas}.  Finally, renormalization is a problem which
has not yet been addressed.
In Fig.~\ref{fig:suscept} we compare the results for
$\chit$ calculated from the eigenvalues of
$M$ and $\gamma_5 M$. For quenched configurations at $\beta=6.0$
both methods give consistent results, indicating that the
neglected renormalization factor is close to one.
In the same figure we compare our results for $\chit$
with those obtained in \cite{Hart:2001fp} using a
gluonic method. The results agree surprisingly well.

\section{CONCLUSIONS}

We calculated the eigenvalues of the Wilson-Dirac operator $M$ and
the Hermitian operator $\gamma_5 M$ on dynamical and quenched
gauge configurations at similar lattice spacings. We found clear
signs for a surpression of small eigenmodes on configurations with
dynamical fermions. We determined the topological susceptibility
using two different definitions of the topological charge based on
the index theorem and a chiral Ward identity. We found good
agreement for quenched gauge configurations at $\beta=6.0$ 
between the two methods.
Our results indicate that the index theorem cannot hold on
coarse lattices since the topological charge turns out to be ill-defined.
Finally, we found a surprisingly good agreement between results for the
topological susceptibility determined with gluonic and fermionic methods.

\vspace*{-0.2cm}
\section*{ACKNOWLEDGEMENTS}

The numerical calculations were performed on the Hitachi {SR8000} at
LRZ (Munich), the Cray {T3E} at EPCC (Edinburgh) and ZIB (Berlin).
We wish to thank all institutions for their support.

\end{document}